\documentclass[12pt,preprint]{aastex}

\usepackage{natbib}

\slugcomment{Draft from \today, to appear in ApJ}

\shorttitle{CN and CS observation in massive star formation}
\shortauthors{Beuther et al.}

\begin{document}

\title{High-spatial-resolution CN and CS observation of two regions of
massive star formation}


\author{H. Beuther}
\affil{Harvard-Smithsonian Center for Astrophysics, 60 Garden Street, 
       Cambridge, MA 02138}
\email{hbeuther@cfa.harvard.edu}
\author{P. Schilke, F. Wyrowski}
\affil{Max-Planck-Institute for Radioastronomy, Auf dem H\"ugel 69, 53121 Bonn, Germany}

\begin{abstract}
Molecular line CN, CS and mm continuum observations of two
intermediate- to high-mass star-forming regions~-- IRAS\,20293+3952
and IRAS\,19410+2336~-- obtained with the Plateau de Bure Interferometer at
high spatial resolution reveal interesting characteristics of the gas
and dust emission. In spite of the expectation that the CN and CS
morphology might closely follow the dense gas traced by the dust
continuum, both molecules avoid the most central cores. Comparing the
relative line strengths of various CN hyperfine components, this
appears not to be an opacity effect but to be due to chemical and
physical effects. The CN data also indicate enhanced emission toward
the different molecular outflows in the region. Regarding CS, avoiding
the central cores can be due to high optical depth, but the data also
show that the CS emission is nearly always associated with the
outflows of the region. Therefore, neither CS nor CN appear well
suited for dense gas and disk studies in these two sources, and we
recommend the use of different molecules for future massive disk
studies. An analysis of the 1 and 3\,mm continuum fluxes toward
IRAS\,20293+3952 reveals that the dust opacity index $\beta$ is lower
than the canonical value of 2. Tentatively, we identify a decreasing
gradient of $\beta$ from the edge of the core to the core center. This
could be due to increasing optical depth toward the core center
and/or grain growth within the densest cores and potential central
disks. We detect 3\,mm continuum emission toward the collimated
outflow emanating from IRAS\,20293+3952.  The spectral index of
$\alpha \sim 0.8$ in this region is consistent with standard models for
collimated ionized winds.
\end{abstract}


\section{Introduction}

Massive star-forming regions consist of hundreds to thousands of
M$_{\odot}$ of dense gas from which future clusters are forming. The
local densities and chemistries vary significantly throughout the
regions, and if one wants to observe the densest regions, and in
particular potential disks at the core center, the right molecular
transitions need to be chosen to study the gas in detail. Especially
the question of disks is still open in spite of large efforts over
recent years. Pure continuum observations lack the necessary velocity
information to prove real disk emission, maser emission often only
allows ambiguous interpretations, and thermal line emission of dense
gas tracers is often contaminated by emission from the ambient gas
(e.g.,
\citealt{cesaroni1997,norris1998,walsh1998,zhang2002,sandell2003}).
Nevertheless, we think that observing thermal emission from the right
molecules~-- which trace disk-like structures and are weak in ambient
gas emission~-- is the most promising path to observe and understand
potential genuine disk emission in high-mass star formation (e.g.,
\citealt{cesaroni1997,zhang2002,beltran2004}). The question remains:
which molecular transitions are best suited to pursue this case?

Many single-dish molecular-line studies of massive star-forming
regions have been undertaken (e.g.,
\citealt{schilke1997b,hatchell1998b,thompson2003}), but only very few
regions have been observed at the high spatial resolution necessary to
disentangle the densest molecular clumps from the potential disks
\citep{cesaroni1999,sandell2003}. Here, we present
high-spatial-resolution Plateau de Bure Interferometer
(PdBI\footnote{IRAM is supported by INSU/CNRS (France), MPG (Germany),
and IGN (Spain).})  observations of two intermediate to high-mass
star-forming regions in two different potential dense-gas and
disk-tracing molecules, CN and CS.

Observations toward low-mass T Tauri stars have shown that CN is
enhanced in their disks \citep{dutrey1997}, and theoretical work by
\citet{aikawa2001} suggested that CN could be a good tracer of gaseous
disks in deeply embedded sources as well. Therefore, we observed the
intermediate-mass region IRAS\,20293+3952 with the PdBI in several
different transitions of CN in the 1 and the 3\,mm band to test the
suitability of CN as a potential massive disk tracer. The mm sources
we are interested in are part of an intermediate-mass sub-cluster in
the near vicinity of an ultracompact H{\sc ii} region which provides
most of the luminosity within the region. For more details on the
source and its multiple outflow properties we refer to
\citet{beuther2004d} and Table \ref{obs_par}.

The other PdBI observations aimed at the high-mass star-forming region
IRAS\,19410+2336 in two CS transitions and the continuum at 1 and
3\,mm. CS has high critical densities ($>10^5$\,cm$^{-3}$) and is
expected to follow the dust emission well \citep{shirley2003}. These
observations were meant to study the large scale CS gas and velocity
distribution as well as tracing the innermost cores and possible
disks. IRAS\,19410+2336 has been studied intensively over recent
years, and the complex outflow and mm continuum properties are
outlined in \citet{beuther2004d} and \citet{beuther2004c}. Further
source parameters are listed in Table \ref{obs_par}.

Section 2 describes the technical details of the observations, and in
section 3 we present the results obtained for both sources. Finally,
section 4 discusses these results in the framework of future dense gas
and disk studies in high-mass star formation.

\section{Observations}

We observed IRAS\,20293+3952 and IRAS\,19410+2336 in several runs
over the two winters 2001/2002 and 2002/2003 with the Plateau de Bure
Interferometer (PdBI) at 1 and 3\,mm. The observing parameters for the
various observations are listed in Table \ref{obs_par}.

The typical system temperature at 1.3\,mm was about 300\,K and at
3\,mm about 135\,K. The phase noise at both wavelengths was mostly
below 20$^{\circ}$ and always below 30$^{\circ}$. Atmospheric phase
correction based on the 1.3\,mm total power was applied. For continuum
measurements, we placed two 320\,MHz correlator units in each band to
cover the largest possible bandwidths. The final continuum images were
produced using only the line-free parts of the spectrum. We estimate
the final flux density accuracy to be $\sim 15\%$. Synthesized beams
at the different wavelengths are listed in Table \ref{obs_par}.

\section{Results}

One of the goals of both observations~-- CN in IRAS\,20293+3952 and CS
in IRAS\,19410+2336~-- was to trace the densest gas and in particular
the massive disks at the very centers of the evolving
protoclusters. Before going into details, we note that at the highest
spatial resolution, in both cases CN and CS do not trace the central
condensations, but mainly gas in their near vicinity. Thus, the data
show that CN and CS are not disk tracers in these two sources. This is
distinctively different from the observations of low-mass T Tauri
disks \citep{dutrey1997}.

\subsection{Millimeter continuum emission in IRAS\,20293+3952}
\label{20293_cont}

The three mm peaks (mm1 to mm3) observed with the PdBI at 2.6\,mm
(Figs. \ref{cn10}, \ref{continuum} ) correspond well with the three
peaks previously observed by \citet{beuther2004d} at 1.3\,mm. The only
differences are that the previous 1.3\,mm data showed a small
sub-component north-east of the main peak mm1, whereas this new data
show an elongation to the south-west of it (mm1a,
Fig. \ref{continuum}). The north-eastern 1.3\,mm sub-component of the
previous dataset could be due to dust emission which might be filtered
out by the larger configurations used to observe the source this time
(AB of PdBI compared to CD in the previous observation). Furthermore,
at 3\,mm the sensitivity for dust emission decreases substantially
($S\propto \nu^{2+\beta}$ with $\beta$ being the dust opacity index),
and therefore we simply might not detect this sub-component at the
lower frequency.

Contrary to this, the sub-component mm1a and its elongation in this new
dataset is spatially correlated with the collimated outflow in this
direction (arrows in Fig. \ref{cn10}, \citealt{beuther2004d}). This is
similar to the recent observations of L1157 where \citet{gueth2003}
distinguish dust emission due to the outflow from the protostellar
dust emission. It is likely that the sub-component mm1a is caused by the
collimated outflow. We will discuss the scenario of a collimated
ionized wind with low spectral index in \S \ref{continuum_discussion}.

Assuming optically thin dust emission for the mm continuum emission,
we can estimate the dust masses of mm1 to mm3. Setting the dust
opacity index $\beta$ to 1 (see the following paragraph) and using a
dust temperature of 56\,K \citep{sridha}, the derived core masses from
the 3\,mm data are between 2.8 and 0.4\,M$_{\odot}$ (Table
\ref{fluxes}), close to the values derived previously by
\citet{beuther2004d}. For mm1 we can also derive a mass from the
1.3\,mm data which is about a factor 2 lower. This difference can be
attributed to the stronger spatial filtering effects of the
interferometer at shorter wavelengths.

{\bf Spectral indices:} The high spatial resolution and good
signal-to-noise ratio of the continuum data allow us to analyze the
spectral index in the mm regime.  However, due to the different
wavelengths, the uv-coverages and thus spatial resolution of the
datasets are different (Fig. \ref{continuum}). To accurately compare
the continuum data it is crucial to employ the same similarly sampled
uv-range (in k$\lambda$) in both bands. Therefore, we flagged all data
outside the uv-range $13-150$\,k$\lambda$ (covered by both frequency
bands) and re-imaged both datasets with the same beam of
$1.5''$. These new datasets are now well suited to determine spectral
indices within the central mm core.

The right panel in Figure \ref{continuum} presents the spectral index
map of the central core mm1 assuming a power law $S\propto
\nu^{\alpha}$ in the Rayleigh-Jeans regime with $\alpha=2+\beta$ where
$\beta$ is the dust opacity index. Since the calibration uncertainty is
15\%, we estimate the error in $\alpha$ and $\beta$ to be $\sim 0.3$.
We find that the spectral index $\alpha$ decreases from $\sim 3.4$ at
the outer edge of the core to values as low as $\sim 2.1$ at the very
center of mm1. Assuming a constant temperature throughout the core,
this corresponds to dust opacity indices $\beta$ decreasing from 1.4
at the core edge to 0.1 at the center. The variation of $\alpha$ is
more pronounced in north-west south-eastern direction perpendicular to
the outflow than in the direction of the outflow. It is likely that
the outflow affects the mm continuum emission of the core in such a
way that the observed $\alpha$ in the outflow direction is lower than
that perpendicular to it (see also the end of this section and
\ref{continuum_discussion}).

Figure \ref{uvfit} outlines a different approach analyzing the data
directly in the uv-plane. The 1 and 3\,mm data can both be fitted with
a Gaussian plus a point source (excluding the data-points $<
20$\,k$\lambda$). This way, we derive spectral indices of the Gaussian
and point source components of 2.7 and 1.6, respectively. Assuming
that the Gaussian component is from a dust core with constant
temperature this results in a dust opacity index $\beta$ of 0.7.  The
central point source could be either a small optically thick disk or
an unresolved hypercompact H{\sc ii} region.

In addition to the main core mm1, we can estimate the spectral index
$\alpha$ for the outflow associated emission feature (Table
\ref{fluxes}). We find a lower spectral index toward mm1a with
$\alpha(\rm{mm1a})\sim 0.8$. Thus, $\alpha$ is lower in the
outflow region than within the central core.

\subsection{CN in IRAS\,20293+3952}
\label{20293_cn}

Figure \ref{cn10} presents the CN(1--0) and 2.6\,mm continuum emission
as contours overlaid on the single-dish large-scale 1.2\,mm image
obtained with the MAMBO-array \citep{beuther2002a}. The small offset
between the single-dish and the interferometric mm peaks is likely due
to pointing inaccuracy of the single-dish data estimated to be $5''$
\citep{beuther2002a}.

On large scales, the CN(1--0) emission follows the mm dust emission
outlined by the MAMBO 1.2\,mm data. But at smaller spatial scales we
find that the CN peaks are always offset from the mm peaks. Because
the interferometric CN line and mm continuum data are taken
simultaneously, the relative positional accuracy between the various
maps is extremely high and the offsets are real. We observe similar
morphologies in the 1.3\,mm band (Fig. \ref{cn21}). It appears that
the CN(2--1) emission forms nearly a ring-like structure around the
continuum source. While in the 1.3\,mm band only one line is strong
enough to be imaged in these observations, in the 2.6\,mm band we have
the main line plus 6 satellite lines due to the hyperfine splitting of
CN (e.g., \citealt{simon1997}). Figure \ref{spectrum_cn10} shows a
spectrum of the main line and one hyperfine component (integrated over
a region of $\Delta \rm{R.A. }[-5,1]''$ and $\Delta
\rm{Dec. }[-5,1]''$). The ratio of the peak intensities of the two
lines is of the order of the theoretical value of $\sim 2.7$
\citep{simon1997}. To first order, this indicates that all lines are
optically thin and trace the whole CN column densities. Therefore, the
weak CN emission toward the mm peaks is a real physical effect and not
due to the optical depth of the observed lines.

One would also like to derive excitation temperature and column
density estimates from the CN(2--1) and (1--0) transitions.  However,
missing flux strongly deteriorating the temperature and column density
estimates is a problem. Therefore, we refrain from further analysis of
physical parameters analysis based on this data. Short spacing
information is crucial to proceed on this path.

\subsection{CS in IRAS\,19410+2336}

The mm continuum data and their interpretation are presented in a
separate paper \citep{beuther2004c}. Although we filter out about
80-90\% percent of the continuum emission, the spatial filtering is
even more severe for the line emission. Comparing the interferometric
measured fluxes (see example spectra in Figure \ref{spectra_cs}) with
the values derived from single-dish observations \citep{beuther2002a},
we find that of the order of only a few percent of the total CS line
emission is recovered in the interferometer data. Therefore, the dust
continuum emission appears to be more strongly associated with the
dense cores than CS.

Figure \ref{cs21} (left panel) presents the integrated PdBI CS(2--1)
observations ($\Delta v$ [17,27] km/s) as a contours overlaid on the
single-dish 1.2\,mm dust emission \citep{beuther2002a}. To associate
the CS emission with the many outflows in the regions, we show with
arrows, ellipses and markers the 7 (or even 9?) outflows in the
region, as presented in \citet{beuther2003a}. Additionally, the right
panel of Figure \ref{cs21} presents the shocked H$_2$ emission in
IRAS\,19410+2336 (also taken from \citealt{beuther2003a}). We find
many different CS emission features throughout the regions (CS1 to CS9
in Fig. \ref{cs21}). However, the overall morphology differs
significantly from the single-dish dust continuum data.  The CS
emission peaks fall into two classes, those in the near vicinity of
the mm peaks (triangles in Figure \ref{cs21}; CS1, CS6 and CS7), and
the rest of the emission peaks which are all offset from the main mm cores.
Comparing these offset CS peaks with the known outflows in the region,
it appears that they are all associated with one or other outflow
feature: CS2 is at the wake of outflow (B), CS3 and CS4 appear to be
associated with the outflows (G) and (H), CS8 is spatially associated
with outflow (J) and also with a strong H$_2$ feature. Finally, CS5
and CS9 both seem to be associated with outflow (F). While CS5 is
again at the wake of the outflow, CS9 is within the outflow lobe where
there is only very weak CO outflow emission \citep{beuther2003a}.

Zooming into the main peak regions in the south
(Fig. \ref{cs54_south}) and in the north (Fig. \ref{cs54_north}, left)
with the higher resolution of the CS(5--4) data, we find that,
although there is emission in the near vicinity of the mm peaks, the
CS emission is offset from the main peaks. As pointed out in \S
\ref{20293_cn}, the positional accuracy between the various maps is
extremely reliable because of the simultaneous interferometric
observations of the CS line and the mm continuum emission.  While we
find emission at all velocities in the north, the southern CS(5--4)
emission is mainly found at blue velocities and the velocity of rest
($\Delta v$ [14,22]\,km/s). There is barely any red CS(5--4) emission
in the southern core at all (Fig. \ref{spectra_cs}). Comparing it
again with the outflow emission \citep{beuther2003a}, we see that the
blue outflow wing (C) follows the same direction as the blue CS
emission. The CS spectra observed toward CS1 (Fig. \ref{spectra_cs})
somewhat resemble the infall spectra observed in CS toward the
massive star-forming region W51 (Fig. 8 in
\citealt{zhang1998b}). However, as our data suffer strongly from
missing flux problems and the negative dip at 23\,km/s is observed
toward positions offset from the peak CS1 as well, we attribute this
feature to observational artifacts due to missing short spacings and
refrain from a physical interpretation.

The image is less clear in the north, the CS peaks are offset from the
mm peak emission (Fig. \ref{cs54_north}), and three outflows emanate
from that region. Nevertheless we cannot unambiguously attribute the
CS emission to any of these three outflows. So, it is very possible
that other reasons contribute to the CS-mm offsets in the north as
well, e.g., optical depth and/or freeze out of CS onto grains in cold
core centers (see, e.g., \citealt{bergin1997,tafalla2004}). In order
to better understand the differences between the CS(5--4) and (2--1)
transitions, we confined both datasets to the same uv-range and
re-imaged these data. The missing flux problems are again severe (see
\S \ref{20293_cn}) but we could produce integrated CS maps of both
transitions (Fig. \ref{cs54_north}, right). Both transitions peak
offset from the continuum, but it appears that the (2--1) transition
is even further offset from the mm continuum peak than the (5--4)
transitions. Different explanations for these observations are
possible: it could be an opacity effect with CS(5--4) turning
optically thick closer to the center than the CS(2--1)
observations. However, physical reasons like higher densities and/or
temperatures toward the core-center can produce this observational
effect as well. Unfortunately, based on this data we cannot
differentiate between the different scenarios~-- independent
observations of temperature and density tracers as well as of
optically thin isotopomers are necessary to get a hand on the
different processes taking place in this region.

Similarly as in \S \ref{20293_cn}, we refrain from temperature and
column density estimates based on the CS data because the missing flux
deteriorates every such an estimate too strongly.

\section{Discussion}
\label{discussion}

\subsection{Line emission}

The search for and investigation of disks in massive star
formation still remains one of the missing links for a more
detailed understanding of high-mass star formation. Theories predict
disks in massive star formation (e.g., \citealt{yorke2002}), and other
observations claim indirectly that disks need to exist because more
and more collimated massive outflows have been found (e.g.,
\citealt{beuther2002d}). However, direct disk evidence in high-mass
star formation is very scarce \citep{cesaroni1997,zhang2002} and yet
not completely unambiguous. Therefore, it is necessary to establish a
procedure to study the dense central cores and potential
disks in massive star formation. We believe that thermal molecular
line emission from dense gas tracers which are not prominent in the
ambient medium is the most promising approach. 

One of the results of this paper is that CS and CN are not well suited
for disk and dense gas studies in these sources. In spite of being a
rather negative result, it needs to be stressed and we recommend the
use of different molecules for future massive disk studies.

High optical depth probably plays a role for CS avoiding the dense
centers of the regions. It remains to be tested whether rarer
isotopomers like C$^{34}$S or $^{13}$CS might trace the dense core
centers and potential disks in a better way. In addition, the
observations demonstrate that the CS emission is dominated to a large
degree by the various outflows in IRAS\,19410+2336. This CS emission
is still associated with the large scale dust emission observed with
single-dish instruments which likely reflects the similar range of gas
densities needed to produce dust continuum and CS line emission.

The situation in slightly different regarding the CN emission in
IRAS\,20393+3952: it also avoids the core centers but optical depth
effects appear to be negligible. UV radiation is necessary to produce
significant CN emission (e.g., \citealt{simon1997}), and CN is
detected toward low-mass disks due to photo-dissociation by the star
on the surface of the disk \citep{dutrey1997,aikawa2001}. Two main
reasons are likely to cause the lack of CN emission toward
IRAS\,20293+3952: (a) the source is too deeply embedded to excite the
surface layers of the potential disk, or (b) the source is in such an
early evolutionary stage that it does not produce enough photons. With
the data available so far, we cannot properly distinguish between the
two possibilities.

These observations outline the difficulties of studying potential
disks in massive star formation. We have to look for other molecular
transitions to properly investigate massive disks.  While NH$_3$
sometimes does a good job \citep{zhang2002} it is often confused by
the ambient gas. The sulphur-bearing species SO was observed toward
the central driving source of HH\,80-81 \citep{gomez2003} but it is
also found toward the outflows in massive star-forming regions (e.g.,
IRAS\,18089-1732, \citealt{beuther2004a}) implying that SO can be
ambiguous as well. The most promising approach might be to focus on
real hot-core-tracing molecules like CH$_3$CN in IRAS\,20126+4104
\citep{cesaroni1997,cesaroni1999}, or HCOOCH$_3$ observed toward
IRAS\,18089$-$1732 with the Submillimeter Array (SMA)
\citep{beuther2004b}. From an observational point of view, the SMA
with a broad bandwidth of 2\,GHz in two sidebands (4\,GHz of
spectral data) offers a unique opportunity to observe many
interesting lines simultaneously and thus compare different molecular
transitions within the same observation \citep{ho2004}. This way, we
should soon get a better census on the disk-tracing molecules in
massive star formation.

\subsection{Continuum emission}
\label{continuum_discussion}

The dust opacity indices $\beta$ we derive from the imaging as well as
from uv-fits are significantly lower than the canonical value of 2
\citep{hildebrand1983}. These results resemble similar recent findings
of low $\beta$ in other massive star-forming regions
\citep{beuther2004b,kumar2003}. However, due to insufficient
signal-to-noise ratio and uv-coverage, the previous study toward
IRAS\,18089$-$1732 was only capable of determining decreasing $\beta$
with increasing baseline length \citep{beuther2004b}. Thus, their data
could not tell whether the decreasing $\beta$ is due to unresolved
clumpy structure throughout the core or an actual decrease in $\beta$
toward the core center. The new data toward IRAS\,20293+3952 now
resolve different spectral indices for the outer regions as well as
the core center. It is not yet clear whether we simply have two
different components~-- a dust core with constant $\beta$ plus either
an optically thick disk or an unresolved hypercompact H{\sc ii}
region~-- or a gradient of varying $\beta$ from the outside to the
core center. Maybe both scenarios contribute to the observed spectral
index variation. A gradient in $\beta$ would be similar to results
that have been reported in low-mass systems by
\citet{hogerheijde2000}. Such a decrease of $\beta$ could either be
due to increasing opacity toward the core center or grain growth
within circum-protostellar disks (see also
\citealt{beckwith2000}). Future work in that field would be required
to cover an even broader range of spatial scales, and it will be
interesting to determine at what distance from the core center $\beta$
reaches the canonical value of 2, if it reaches it at all.

The finding that $\alpha$ is lower toward the outflow than toward the
core center is different from the recent results from \citet{gueth2003}
in the low-mass outflow L1157, who reported an average spectral index
of $\alpha \sim 5$ toward the outflow. Their analysis is based on
bolometer data over very broad band-passes ($\sim 80$\,GHz) and thus
line contamination can introduce artificial flux increases in one or
both bands. But even taking such artifacts into account,
\citet{gueth2003} conclude that in L1157 $\alpha$ is higher toward the
outflow than toward the core. Their observations point toward smaller
dust grains and/or a different chemistry, which is consistent with
models predicting the evaporation of icy dust grains during shock
interactions of the outflow with the surrounding dust.

These models obviously fail toward mm1a. However, thermal ionized
wind models predict spectral indices between 2 and $-0.1$
\citep{reynolds1986}. A typical spherical wind results in $\alpha
\sim 0.6$ whereas highly collimated jets have even lower values around
0.25, and in the optically thin case $-0.1$
\citep{reynolds1986}. Therefore, the derived $\alpha(\rm{mm1a})\sim
0.8$ within the outflow region of IRAS\,20393+3952 is consistent with
standard models of collimated ionized winds.

\acknowledgments{We very much appreciate the comments from an
anonymous referee which helped to improve the manuscript. We would
like to thank the staff of IRAM in Grenoble for help during the
observations and the data reduction process. H.B. acknowledges
financial support by the Emmy-Noether-Program of the Deutsche
Forschungsgemeinschaft (DFG, grant BE2578/1).}


\clearpage
\begin{deluxetable}{lrr}
\tablecaption{Observation parameters \label{obs_par}}
\tablewidth{0pt}
\tablehead{
\colhead{} & \colhead{IRAS\,20293+3952} & \colhead{IRAS\,19410+2336}
}
\startdata
Date       & winter 2002/2003 & winter 2001/2002 \& 2003/2004 \\
R.A.[J2000]& 20:31:12.9         & 19:43:11.4  \\
Dec.[J2000]& +40:03:22.9        & +23:44:06 \\
$v_{\rm{lsr}}$ & 6.3\,km/s      & 22.4\,km/s \\
Distance$^a$ [kpc] & 2.0            & 2.0   \\
Config.    & AB           & BCD \\
Lines      & CN N=1--0 \& N=2--1 & CS J=5--4 \& J=2--1 \\
Freq.[GHz] & 113.385 \& 226.656 & 97.981 \& 244.936 \\
Freq. res. & 0.312\,MHz         & 0.039\,MHz \\
Velo. res. & 0.83 \& 0.41\,km/s & 0.12 \& 0.05\,km/s \\
$\theta (1\,\rm{mm})$ & $0.76''\times 0.61''$(P.A.$37^{\circ}$) &
                        $1.52''\times 1.03''$(P.A.$63 ^{\circ}$) \\
$\theta (3\,\rm{mm})$ & $1.46''\times 1.21''$(P.A.$32^{\circ}$) &
                        $5.31''\times 3.42''$(P.A.$77 ^{\circ}$)\\ 
Mosaic offsets [$''$] &   ---  & -2/-2 -12/-8 9/2 -1.5/9.5 \\
                      &   ---  & -9.5/55 -19.5/49 -9.5/43 \\ 
                      &   ---  & -19.5/37 -19.5/25 -10/17\\ 
Phase calib.& 2013+370, 2023+336& 2023+336, 1923+210 \\
Amp. calib.&  MWC349            & MWC349 \\
$3\sigma$\,rms(1.3mm) & 3.6\,mJy & 9\,mJy$^b$ \\
$3\sigma$\,rms(2.6mm) & 1.2\,mJy & 0.75\,mJy$^b$ \\
\enddata
\tablenotetext{a}{\footnotesize The kinematic distance ambiguities
are solved by associating the regions via the near- and mid-infrared
surveys 2MASS and MSX on larger scales with sources of known distance
(Bontemps, priv. comm.).}
\tablenotetext{b}{\footnotesize Data presented in \citet{beuther2004c}}
\end{deluxetable}

\begin{deluxetable}{lrrrrrrrrrr}
\tablecaption{Continuum data in IRAS\,20293+3952 \label{fluxes}}
\tablewidth{0pt}
\tablehead{
\colhead{Source} & \colhead{Offset} & $S_{2.6\rm{mm}}^{\rm{int}}$ & $S_{2.6\rm{mm}}^{\rm{peak}}$ & $S_{1.3\rm{mm}}^{\rm{int}}$ & $S_{1.3\rm{mm}}^{\rm{peak}}$ & $\alpha$ & $M_{3\rm{mm}}$ & $M_{1\rm{mm}}$ \\
\colhead{} & \colhead{[$''$]} & [mJy] & [mJy/beam] & [mJy] & [mJy/beam] & & [M$_{\odot}$] & [M$_{\odot}$]\\
}
\startdata
mm1  & 0/0         & 28.0& 16.1 & 118.0 & 42.3 & $<3.4^b$ & 2.8 & 1.6 \\
mm1a & $-2.7/-1.3$ & 2.9$^a$ & 2.9 & 4.8 & 4.8 & $\sim 0.8$ & -- & -- \\
mm2 & $-10.2/-10.7$& 8.6 & 5.6 & -- & -- & --& 0.9 & -- \\
mm3 & $-5.6/-12.4$ & 3.6$^a$ & 3.6 & --& -- & -- & 0.4 & -- \\
\enddata
\tablenotetext{a}{\footnotesize Source is unresolved.}
\tablenotetext{b}{\footnotesize See the main text.}
\end{deluxetable}

\clearpage
\begin{figure}[htb]
\caption{IRAS\,20293+3952: the grey-scale shows the 1.2\,mm
large-scale emission obtained with the MAMBO-array
\citep{beuther2002a}, the grey-scale levels are 18 to
336\,mJy/($11''$beam) in 36\,mJy/($11''$beam) steps
(18(36)336\,mJy/($11''$beam)). The contours present the PdBI
observations: white contours the 2.6\,mm continuum
(1.2(1.2)15.6\,mJy/beam), and black contours the integrated CN(1--0)
emission ($\Delta v$ [4,12]\,km/s, 29(42)407\,mJy/beam). The arrows
sketch the four outflows previously observed by \citet{beuther2004a},
and the beam is shown at the bottom left. The small box outlines the
region presented in Figure \ref{cn21}.}
\label{cn10}
\end{figure}

\begin{figure}[htb]
\caption{Millimeter Continuum emission in IRAS\,20293+3952. The left
and middle panels show the 2.6\,mm and 1.3\,mm emission,
respectively. The contouring is done in $3\sigma$ steps
(1.2\,mJy and 3.6\,mJy, respectively).  The right panel presents in
grey-scale the spectral index map between the two bands and in
contours the 1.3\,mm continuum emission. For the right panel, the data
of both bands were selected to cover the same uv-range (in
k$\lambda$).}
\label{continuum} 
\end{figure}

\begin{figure}[htb]
\caption{The 1.3 and 2.6\,mm continuum data in the uv-plane and the
corresponding two-component fits. The top data-points (triangles) are
the 1.3\,mm data and the bottom data-points are the measurements at
3\,mm (squares). The dotted lines show the fitted point source and the
solid lines the combined fits consisting of a Gaussian plus a point
source.}
\label{uvfit}
\end{figure}

\begin{figure}[htb]
\caption{IRAS\,20293+3952: the grey-scale with dotted contours shows
the 1.3\,mm continuum emission (3.6(3.6)39.6\,mJy/beam), and the lined
contours present the integrated CN(2--1) data ($\Delta v$ [4,9]\,km/s,
111(74)703\,mJy/beam).  The beam is shown at the bottom right.}
\label{cn21}
\end{figure}

\begin{figure}[htb]
\caption{CN(1--0) spectrum in IRAS\,20293+3952 integrated over a
region of $\Delta \rm{R.A. }[-5,1]''$ and $\Delta \rm{Dec. }[-5,1]''$,
the two different transitions are labeled at the top.}
\label{spectrum_cn10}
\end{figure}

 
\begin{figure}[htb]
\caption{IRAS\,19410+2336: CS(2--1) and CS(5--4) spectra in full and
dashed lines, respectively. The spectra are taken toward the positions
CS1 (within 6\,arc-sec$^2$) and CS6 (within 9\,arc-sec$^2$)
(Fig. \ref{cs21}). The CS(5--4) spectrum toward CS1 is scaled up by a
factor 2.}
\label{spectra_cs}
\end{figure}

\begin{figure}[htb]
\caption{IRAS\,19410+2336: {\bf Left panel:} the grey-scale shows the
  1.2\,mm large-scale emission obtained with the MAMBO-array
  \citep{beuther2002a}, the grey-scale levels are
  85(85)765\,mJy/($11''$beam). The contours present the integrated
  CS(2--1) PdBI emission ($\Delta v$ [17,27]\,km/s,
  0.2(0.2)1.8\,Jy/beam).  The arrows, ellipses, triangles and labels
  mark the outflows and mm sources as outlined in
  \citet{beuther2003a}. The 3\,mm beam is shown at the bottom
  right. {\bf Right panel:} the grey-scale now presents the infrared
  shocked H$_2$ emission taken from \citet{beuther2003a}. The contours
  again show the integrated CS(2--1) emission, and the numbers label
  the different CS emission features discussed in the main body.}
\label{cs21}
\end{figure}

\begin{figure}[htb]
\caption{IRAS\,19410+2336: the grey-scale shows the 1.3\,mm continuum
data already presented in \citep{beuther2004a}, the grey-scale levels
are 7(7)63\,mJy/beam. The contours outline the blue CS(5--4) emission
in that region ($\Delta v$ [14,19]\,km/s, 0.5(0.5)4.5\,Jy/beam). The
arrows sketch outflow C from \citet{beuther2003a}, the synthesized
beam is shown at the bottom right.}
\label{cs54_south}
\end{figure}

\begin{figure}[htb]
\caption{IRAS\,19410+2336: {\bf (left):} the grey-scale shows the
  1.3\,mm continuum data already presented in \citet{beuther2004a},
  the grey-scale levels are 9(3)50\,mJy/beam.  The contours outline
  the integrated CS(5--4) emission in that region ($\Delta v$
  [17,24]\,km/s, 0.7(0.7)6.3\,Jy/beam). In the {\bf right} image, we
  present the CS and 1.3\,mm continuum data imaged all with the same
  uv-range for better comparisons. The grey-scale shows the continuum,
  full contours CS(5--4) and dashed contours (CS(2--1), the contouring
  for all three is from 5 to 95\% (step 10\%) of their peak emissions,
  respectively. The synthesized beams are drawn at the bottom left of
  each panel.}
\label{cs54_north}
\end{figure}

\end{document}